\documentstyle[12pt,epsfig]{article}
\begin{document}

\begin{flushright}
MPI-PhT/98-144\\
October 1998\\
\end{flushright}
\vspace{20mm}
\begin{center}
\large {\bf Space and Family}\\
\mbox{ }\\
%\vfill
\normalsize
\vskip3cm
{\bf Bodo Lampe}               
\vskip0.3cm
Max Planck Institut f\"ur Physik \\
F\"ohringer Ring 6, D-80805 M\"unchen \\
%%\vspace{20mm}
\vspace{1.5cm}

{\bf Abstract}\\
\end{center}
Geometrical pictures for the family structure of fundamental 
particles are developed. They indicate that there 
might be a relation between the family repetition structure 
and the number of space dimensions.

\newpage

{\bf 1. Introduction} 

With the discovery of the top quark  
all in all 12 fundamental 
fermions (6 leptons and 6 quarks) are known today. 
At least a decade ago it has become clear that these particles 
can be organized into three "families" 
each containing 2 quarks and 2 leptons. 
These particle families (or generations) behave 
identically under the electroweak and strong 
interactions and do not differ by anything else than their 
masses.   
The number of generations will probably be 
restricted to three forever, because it has been shown 
experimentally that at most 3 species of light neutrinos 
exist. A fourth family,if it exists at all, 
would necessarily contain a heavy 
neutrino and would therefore be different in nature 
from the known families.   

Over the last 20 years there has been one outstanding puzzle 
in elementary particle physics. This is 
the question whether the variety of "elementary" particles, 
the quarks and leptons, can be derived from some more 
fundamental principle. To answer this question is quite 
difficult because up to now no experimental indications 
exist of which might be the nature of this principle. 
Present models usually lead to an inflation of new 
particles (like supersymmetry) at higher energies and/or tend to 
shift the basic problems to higher energy scales where they 
reappear in slightly modified form (like technicolor). 

Preon models (e.g. \cite{harari}) avoid these deficiencies but have 
severe problems of other kinds, like the smallness of fermion masses 
as compared to the binding scale. Still, I want to follow in this 
work a preon type idea, 
that the quarks 
and leptons have a spatial extension, and contain most probably 
sub--constitutents. 
My guideline will be that the spatial 
dimensions correspond to a sort of shells which are 
successively filled up by the generations.  
The third shell -- corresponding in some sense to the third dimension --  
becomes closed with the top quark. 
Several "pictures" will be presented which en gross adhere to this 
general philosophy but differ in the details of its realization. 
I shall also address the question of how to understand the 
vector bosons and the mass hierarchy. 

I shall make use of some discrete, nonabelean subgroups of 
O(3), in particular the symmetric group $S_4$. Discrete subgroups 
have been repeatedly studied in the literature to tackle 
the generation problem. However, the 
approach to be followed here is much different. For example, a spatial 
extension of the fermions will be assumed.  

I would like to warn the reader that in this article I am mostly 
doing simple minded geometry without really clarifying 
the mathematics behind it. 
There are all sorts of unanswered questions concerning the 
dynamics of the model. 
%left as for the understanding 
%of the dynamics.  
My hope is that I can motivate readers to do more refined work 
on the basis of these suggestions.

%\begin{figure}
%\begin{center}
%\epsfig{file=spacefig2.eps,height=6.5cm,angle=270}
%\caption{ }
%\label{fig2}
%\end{center}
%\end{figure}

\vskip2cm

{\bf 2. The Family Repetition} 

The known quarks and leptons within one family are: 
\begin{equation} 
 \nu_{L,R} \qquad e_{L,R} \qquad u^{1,2,3}_{L,R} \qquad d^{1,2,3}_{L,R}  
\label{eq1}
\end{equation}
where the upper index denotes the quark color degrees of 
freedom and the lower index the helicity. I have assumed 
here that a righthanded neutrino $\nu_R$ exists, 
because it exists naturally in most of the 
pictures developed below. 
Of course, one can very well 
have parity violation {\it and} righthanded neutrinos just 
by demanding that the observed W and Z only interact with lefthanded 
currents. However, one encounters new heavy vector bosons 
($W_R$ and $Z_R$) to interact with the righthanded neutrino. 

There are $ 2_{spin} \times 2_{isospin} \times (1+3_{color})=16$ 
degrees of freedom in one family (32, if 
antiparticles are counted as well) and altogether 48 
degrees of freedom in all three generations. 
This might be related to a property of 3--dimensional space, 
namely that there are 
48 symmetry transformations which leave the 
3--dimensional cube invariant. Furthermore, 
as will be shown later, there are subgroups 
of this symmetry group which can be arranged in such a way 
that they look like the quark--lepton arrangement in the 
3 families. Although this may be just by accident, I will 
explore in this article the implications 
of such a relation. 
%of an intimate connection between the geometry of the 
%cube and the observed quark--lepton spectrum. 

Any of the following models should have the 
ability to generate the quantum numbers of the observed 
fermions in a correct way. For example, the electric charges 
should come out correctly as multiples of one third 
because this is one of the main requirements which leads one 
to conclude that quarks and leptons are of the same origin. 

It is well--known that up- and down type quarks take part 
in all the standard interactions (strong and 
electroweak) whereas the leptons 
do not couple strongly, and the neutrinos couple only weakly. 
These facts are reflected by the quantum numbers;  
neutrinos carry only the (weak) isospin, electrons 
an electric charge in addition, and  
quarks carry a color charge, an electric charge 
and a nontrivial weak isospin quantum number. 
%\footnote{Since weak and electromagnetic  
%interactions are mixed, it is sometimes more convenient 
%to use the weak hypercharge $Y=Q+{I_3 \over 2}$ instead of Q.  
%(Q and $I_3$ are electric charge and weak isospin, 
%respectively.)}  
The fact that neutrinos interact only weakly, is probably 
related to their tiny masses, and indicates 
that their "sub--shells" are somewhat more closed than those of 
the other fermions. Similarly, since no lepton interacts 
strongly, they should be considered more saturated bound states  
than the quarks.  

One might think that some information on the nature of the 
fermions can be obtained from their measured mass spectrum.  
However, the fermion masses are running, i.e. energy dependent, 
and it is not really 
known which dynamics governs this energy dependence.  
In other words, their enormalization group 
equations are not known precisely and therefore no complete 
knowledge of the fermion mass spectrum exists. 
For instance, the masses could be running according to some SUSY--GUT 
theory. However, apart from the fact that the SUSY breaking scale 
is not known precisely, new physics  
may set in at some point and modify the RG equations. 
Therefore it is not known to what values the fermion masses are 
converging. There might be relations like $m_b=m_{\tau}$ 
at scales $\sim 10^{16}$ GeV, but those  
are not compelling. 

I think it is fair to say that we only have a 
rough knowledge of the fundamental 
mass parameters. The fermion mass spectrum is at most a qualitative 
guideline to understand the family structure. 
 
As compared to the unification scale the masses of the known fermions 
are tiny. However, if compared among themselves 
the mass differences between the families and also within one 
family are so vastly different that one should work 
with mass ratios instead of mass differences to describe them. 
In spite of the above mentioned principle limitations  
one can sort out a few basic masses and ratios which will probably 
survive the RG running. 
Among them there is the "overall mass" of a family, i.e. 
the average mass of its non--neutrino components. One has 
roughly 
\begin{equation}
m_I=10^7 \qquad m_{II}=10^9 \qquad m_{III}=10^{11}  
\label{eq2}
\end{equation}
(in eV) for the first, second and third family, i.e. 
a factor of about 100 between the masses of successive
generations. 
% One can make 
% this rough statement more quantitative by defining the family mass to be 
% $m_F={N_c (m_U +m_D) +m_L \over 7}$ where $N_c$ is the number of 
% colours. Depending on the future results for the top quark mass 
% and on the RG running  
% it may very well turn out that the mass ratios 
% ${m_{III} \over m_{II}}$ and ${m_{II} \over m_{I}}$ are the same. 

Secondly there is the ratio between the neutrino and the 
average family mass $m_F$,  
which may be either zero or 
of the order 
\begin{equation}
{m_{\nu F} \over m_{F}} =10^{-10}  \qquad , F=I,II,III 
\label{eq3}
\end{equation}
according to recent neutrino data 
and its smallness should be qualitatively explained by any model. 

\begin{figure}
\begin{center}
\epsfig{file=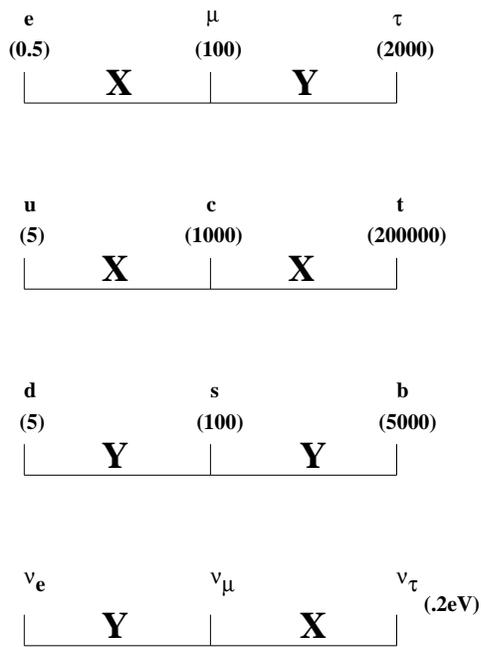,height=6.5cm,angle=270}
\vskip 0.8cm
\caption{Approximate mass ratios X and Y between particles of identical 
quantum numbers in successive generations. The numbers in brackets 
are the approximate fermion masses (in MeV) which lead to these mass 
ratios. The neutrino mass ratios are guessed. }
\label{figmr}
\end{center}
\end{figure}

There are, thirdly, two less reliable mass ratios which I call X and Y 
and which arise if one looks at the approximate mass values of the 
fermions in the three generations. Namely, one realizes that the 
mass ratios 
\begin{equation}
X \equiv {m_{\mu} \over m_{e}} \sim {m_{c} \over m_{u}}  
\sim  {m_{t} \over m_{c}} \sim 200 
\label{eq4}
\end{equation}
are approximately equal. 
The same holds true for 
\begin{equation}
Y \equiv {m_{\tau} \over m_{\mu}} \sim 
{m_{b} \over m_{s}} \sim {m_{s} \over m_{d}} \sim 20 . 
\label{eq5}
\end{equation}
These relations are visualized in fig. \ref{figmr}. 
Admittedly they are very rough and  
could be spoiled by RG running, but they are interesting 
enough to be shown.   
I have included in fig. \ref{figmr} an educated guess concerning 
the neutrino mass ratios. Starting from a mass $m(\nu_\tau)=0.2$eV 
one is lead to $m(\nu_\mu)\approx 0.001$eV and $m(\nu_e)\approx 0.00005$eV 
which is in accord with recent results from solar--neutrino and 
Super--Kamiokande data. 

To fully understand the fermion masses a detailed knowledge of 
the dynamics inside the fermions would be necessary. 
The models to be presented 
in this work are not able to provide this. 
It would already be progress if some qualitative features like 
the smallness of the neutrino 
masses as compared to the other fermions could be explained. 
Furthermore, there is the puzzle that {\it all} 
observed fermion 
masses are much smaller than the scale at which they are bound. 
t'Hooft \cite{thooft} has suggested to decree chiral invariance 
as the principle which suppresses the fermion masses, and from this has  
derived conditions on the anomaly structure of the preon model. 
In the present work this is not a necessary condition. The point 
is that the extension of the observed fermions is not fixed 
by the binding energy of a superstrong force but 
{\it by the structure of space}. For example, in the model 
presented in section 5, space is essentially discrete with preons 
sitting on the sites of a cubic lattice. I shall stick 
to the notion of 'binding energy', though, to mean the inverse extension 
of the 'bound states'. 
Actually, the binding energy is the scale at which the fermion 
masses should be defined.  

For experiments at high energies the precise values of the 
fermion masses become less relevant. The only masses of 
actual significance at medium and high energies $(\geq m_W)$ are 
$m_W$, $m_t$ and $m_H$. For practical purposes all the other 
masses may be put to zero. 
It would be extremely interesting 
if a mass relation of the form
\begin{equation}
{m_H \over m_W}=G({m_t \over m_W}) 
\label{eq6}
\end{equation}
would exist. Unfortunately, the Higgs particle is not a very 
natural object in the models to be presented. It may be 
constructed in some of the models, along similar lines 
as the vector bosons, but its existence is not compelling. 
To guarantee renormalizability of the low energy theory at 
small distances one should probably take it in. 

\vskip2cm

{\bf 3. Basic Assumptions} 

The basic assumption in this work goes as follows:  
the fermions in the first family can be considered as 
effectively one--dimensional objects composed of a 
shell which is successively filled up when one goes from 
the electron--neutrino to the up--quark. 
This is not to say that they are truely one--dimensional, 
but that their structure can be encoded in such a way 
as to correspond to one of the three spatial dimensions. 
We shall have several examples for that below. 
 
The closed first family shell survives in the second family 
where another shell is beginning to fill. The second family 
is thus becoming 2--dimensional in nature. 
Finally, the third family fills 3-dimensional space 
completely. The structure is completed with the top 
quark. 'Completeness' does not necessarily  
correspond to a saturation in the sense 
that the top quark mass would be lowered by the fact that  
the top quark corresponds to a closed shell. On the 
contrary! The particles mostly 'saturated' are certainly 
the neutrinos, because their masses are extremely 
small and they interact only weakly. The models to be  
presented try to take that into account. 

To account for the rather large mass difference 
between the overall family masses (factors $\sim 100$) one may 
speculate that they arise from "exciting" the successive dimensions.  

In most of the models considered below, preons are naturally 
assumed to be massless, or -- less restrictive -- 
have masses much smaller than their binding energy.  
Some of the vastly different masses of fermions may be 
due to preons of different mass, though.  
For example, to account for the extremely large mass difference
between the neutrinos and quarks within one generation 
(a factor of order $10^9$ at least) 
one may  introduce some sort of sub--quark 
present in quarks (and non-neutrino leptons) but not in 
neutrinos. This picture is more in accord with t'Hooft's 
chiral invariance condition, where the difference in the  
observed fermion masses is attributed to the fact that the 
preons have different masses, and not to the (large) binding 
energy. 

At first sight the large mass ratios between the 
families and also within one family make it difficult 
to believe that all particles can be derived from a universal 
symmetry principle. Still, all these masses are probably 
small as compared to the binding energy of the 
approximately massless preons. 
 
In section 4 some qualitative pictures are developed to understand 
the connection between space and family. 
The model to be discussed most extensively will be presented 
afterwards in sections 5 and 6. 
There the existence of families will be tied to the 
3 spatial dimensions, but instead of shells in the narrow sense 
we shall have 4 preons interacting with each other in various 
permutations, the set of all permutations ($S_4$) exhausting the 
fermionic and bosonic degrees of freedom. The preons will be sitting in the 
corners of a tetrad, to form a fermion. A vector boson can 
be obtained by fusion of 2 such tetrads to form a cube.  

\vskip2cm

{\bf 4. Some qualitative Pictures and Guidelines}   

\begin{figure}
\begin{center}
\epsfig{file=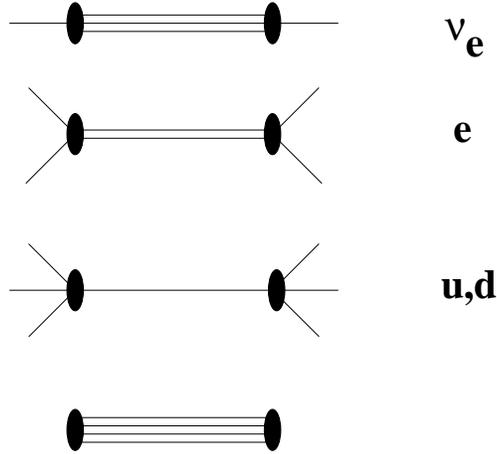,height=6.5cm,angle=270}
\vskip 0.8cm
\caption{a 1--dimensional ordering of the first family fermions.}
\label{fig0}
\end{center}
\end{figure}

In the following I want to develop a simple picture based 
on the 3--dimensional structure of space. 
To form the first generation, only one axis ('x') 
%the first(=x)--axis  
is populated. To build up the second family 
the y--direction is filled up , etc. 
An example of how this may work, is shown in fig. \ref{fig0}.  
The fermions of the first generation are made up from 
two preons, drawn as full dots in fig. \ref{fig0}, 
and bound by forces which can be visualized as three, two or one  
lines connecting the presons. 
The graphs in fig. \ref{fig0} with one, two or three 
lines connecting the presons correspond to $\nu_e$, $e$ and
$u,d$, repectively. 
As another rule we demand that there are exactly 4 lines connecting 
to each preon. As a consequence, $\nu_e$, $e$ and
$u,d$ have two, four and six open, 'unsaturated' 
strings, respectively, cf. fig. \ref{fig0}.

It should be noted that a particle  
exists in this picture, which is completely 
saturated and therefore does not take part in any 
of the known interactions -- apart from, perhaps, 
gravity. This particle is shown  in 
the lower part of fig. \ref{fig0}. If it has a (tiny) 
mass, it would be a 'singlet' type dark matter 
candidate. 

The second generation  
is formed by doing the same construction as done 
for the first, but along the y--direction. The procedure 
for the second family ends 
when the singlet state is formed along the y--direction.
Note that one starts with quarks in fig. \ref{fig0}, and goes
via leptons and neutrinos, to the singlet state.
It will be assumed that the singlet along 
the x--direction is present in all second (and third) family 
fermions, so that the second family is of 2--dimensional nature. 

Finally, the third generation is made up along the 
z--direction, according to the same rules depicted in 
fig. \ref{fig0}. Note that in the third family the 
x-- and y--direction are occupied by the first and 
second family singlet states and that the third 
generation fills 3--dimensional space completely.  

In this picture weak,electromagnetic and strong interactions 
are a question of the number of open strings emerging 
from a preon. The strong interaction occurs 
when three open lines join together to form a bound 
state, for the electromagnetic interaction two lines are needed 
and for the weak interaction only one line of a lepton or 
quark connects with one line of another lepton or quark.  

An important question is how the multiplicity  
of quarks arises. We have one lepton degree of freedom 
for each of the first pictures in fig. \ref{fig0}, but 
six quarks corresponding to the third picture in 
fig. \ref{fig0}. One possibility is to make a rule 
which counts the number of ways the open strings 
of one fermion can be connected. There is one possibility 
for the neutrino ($\nu_L$), two for the other leptons 
($l_{L,R}$) and 6 for the quarks. 
Interaction processes like $\nu \bar{\nu} \rightarrow q \bar{q}$ 
may be understood as rearrangement of open and closed strings. 

Furthermore, there is the question of how the fermion 
masses can, at least qualitatively, be understood. 
One can either try to understand them dynamically, 
from the magnitude of the fermion binding energies.  
Another possibility is to put a third, massive, preon 
into the centre of the electron and quarks, but not 
the neutrino. 

The presented picture is not very sophisticated and  
certainly not complete. It is meant as a warming 
up for the more elaborate model to follow in section 5 and as  
a qualitative guideline of how families can be 
related to three spatial dimensions. 

\begin{figure}
\begin{center}
\epsfig{file=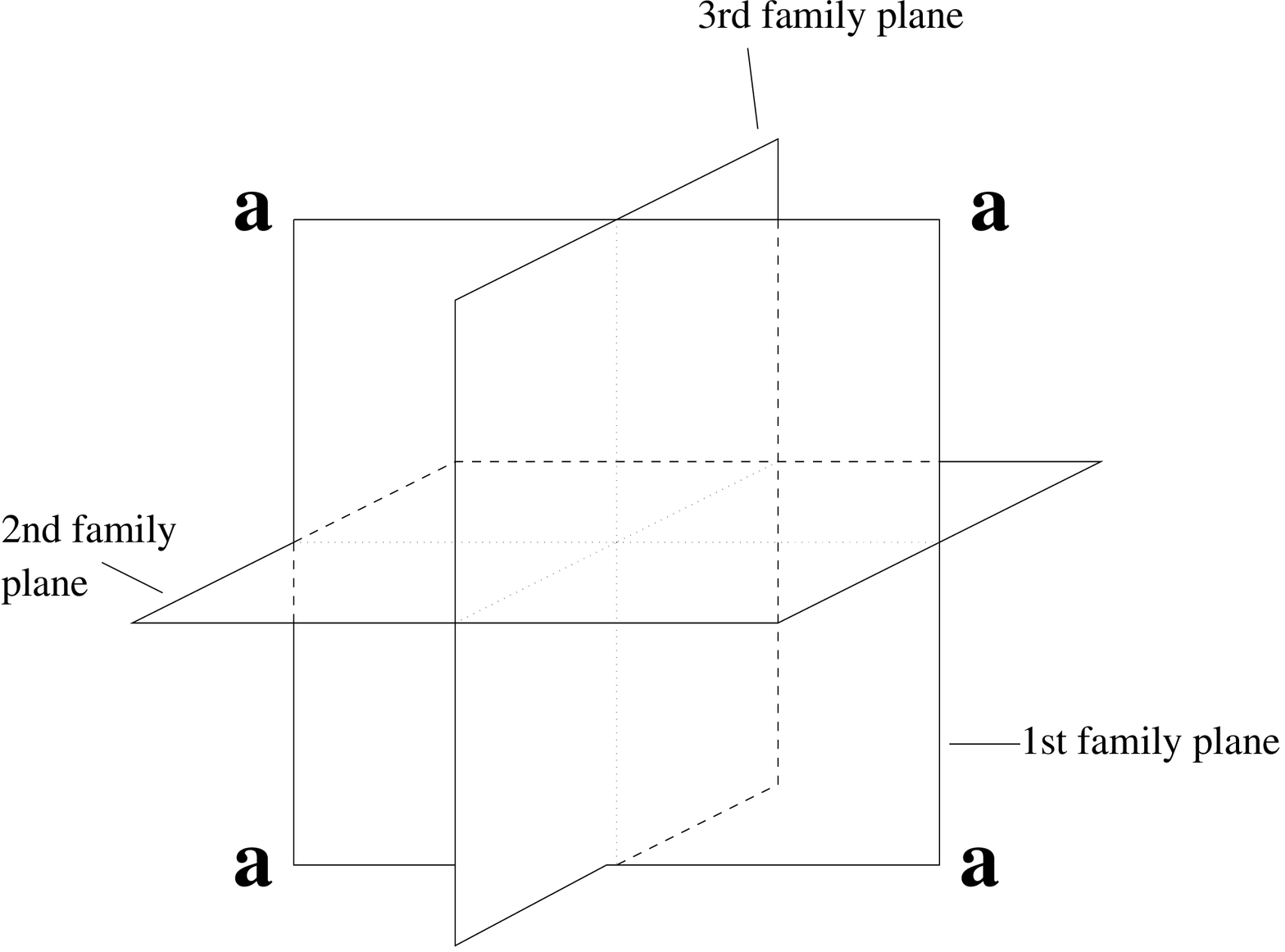,height=9.5cm,angle=270}
\vskip 0.8cm
\caption{ }
\label{fig4}
\end{center}
\end{figure}

There are a lot of variations of the model presented here. 
For example, instead of constructing the fermions of one family 
by aligning the preons along spatial axes, one could 
align them in the 3 planes orthogonal to the coordinate axes,   
cf. fig \ref{fig4}. 
In such a picture one could, for example, put 4 identical 
preons $a$ on the corners of a 
square in the 'first family plane', cf. fig. \ref{fig4},  
and, perhaps, another (massive) preon in the centre of 
the square (except for the neutrinos). 
If $a$ has a quantum number which can take values 
$\pm 1$, one can form the 16 states necessary to build 
up a family. Note that in this picture there would be 
a right handed neutrino.    

Alternatively, one may develop a more dynamical picture 
in which two preons (with certain quantum numbers, to get the 
complete 16plet 
of fermions of the first generation)  
encircle each other in the first family 
plane. As for the second family, there are 2 other preons 
encircling each other in a plane orthogonal to 
the first one, etc. 
%  until finally in the third family one 
% has the situation of 6 particles depicted in fig. \ref{fig2}. 

\begin{figure}
\begin{center}
\epsfig{file=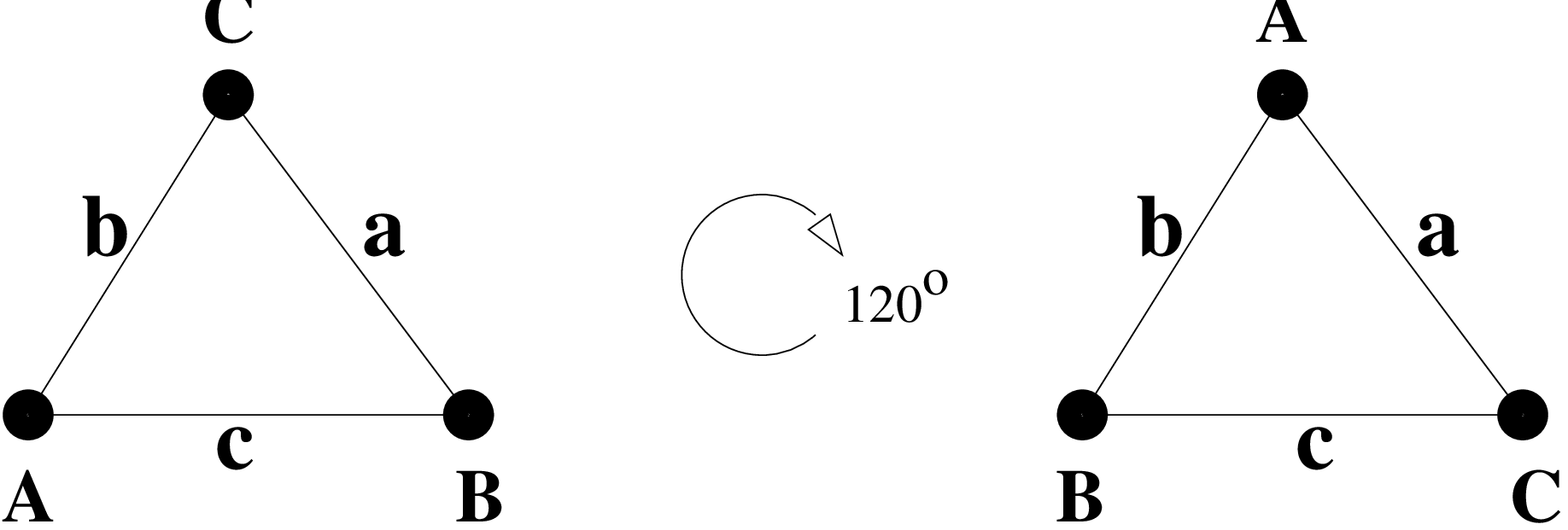,height=9.5cm,angle=270}
\vskip 0.8cm
\caption{ }
\label{figt}
\end{center}
\end{figure}

Clearly, all of these pictures give no proof of the claim 
that the number of generations 
is tied to the number of dimensions. 
There is a more sophisticated picture in section 5 which might 
serve a better job.  
It centres around discrete subgroups of O(3). 
Of particular relevance will be the group of permutations 
$S_4$ which is isomorphic to the symmetry group of the 
tetrad.  
The discussion will concentrate on the subgroups themselves 
and not on their representations. This is a somewhat unusual 
approach, because normally in physics particle multiplets  
are identified with the representation spaces and not with the 
symmetry groups themselves. In contrast, the philosophy 
here is that by applying the symmetry transformations on the 
ground state ($\nu_e$) one can generate all other fermion 
states. This is only possible if there is a symmetry {\it breaking} 
which distinguishes the generated states. Such a symmetry 
breaking will be realized by geometrical means in the following 
section. To show how this can happen, I have visualized 
in fig. \ref{figt} an element of $S_3$, the permutation symmetry 
of the three sides  
of an equilateral triangle in 3 dimensions, assuming the 
existence of distinct preons A,B,C on the sites and distinct 
binding forces on the links of the triangle. If the $S_3$--transformation 
is applied to the sites but not to the links of the triangle, 
a completely different state is generated. In this case there are 6 such 
states corresponding to the 6 elements of $S_3$, which may have 
a relation to the weak vector bosons (see section 5). 

\vskip2cm

{\bf 5. The Model}

The following picture rests on the geometry of squares and cubes.
The idea is to associate a particle degree of freedom to 
each of the symmetry transformations.  
In the case of squares there are 16 such transformations, 
among them the trivial unit operation,  
whereas for cubes there are 48. 

\begin{figure}
\begin{center}
\epsfig{file=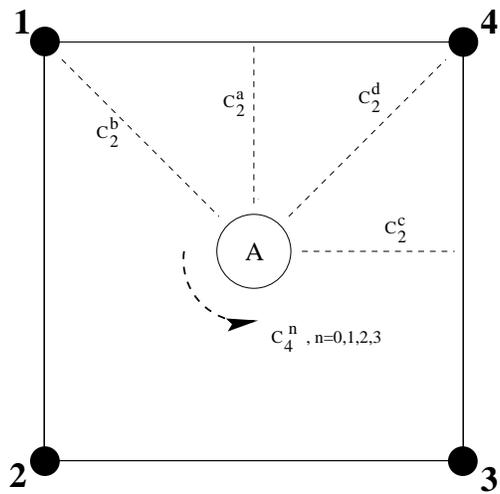,height=6.5cm,angle=270}
\vskip 0.8cm
\caption{The symmetry transformations in $D_4$. }
\label{fig1}
\end{center}
\end{figure}

As a warming up 
consider the set $S_s$ of symmetry transformations of the square 
embedded in 3--dimensional space. This yields a formal 
describtion of a one--family situation, in the following sense:  
It consists of 16 elements and is a direct product of the   
parity transformation $P \equiv (\vec{x} \rightarrow -\vec{x})$ with the 
set $D_4$ of 8 rotations 
depicted in fig. \ref{fig1}, $S_s=P \oplus D_4$. In fig. \ref{fig1},  
$C_n$ stands 
for a rotation by $2\pi /n$ with the rotation axis being indicated. 
The parity transformation is clearly destinated to correspond 
to the two possible helicities of fermions. 
%The subgroup formed 
%by the 8 rotations of fig. \ref{fig1} is isomorphic to the group 
%of permutations $S_3$. das ist falsch!Es gilt Sc=PxS4,aber nicht 
% Ss=PxS3;denn S3 hat nur 6 Elemente.
%Written in the language of permutations 
%we associate the following group elements to quarks and leptons, 
%$\nu \longleftrightarrow (+++)$, 
%$e \longleftrightarrow (---)$, 
%$u^1 \longleftrightarrow (+--)$, $d^1 \longleftrightarrow (-++)$, 
%$u^2 \longleftrightarrow (++-)$, $d^2 \longleftrightarrow (--+)$,
%$u^3 \longleftrightarrow (+-+)$, $d^3 \longleftrightarrow (-+-)$. 
%dies ist nun wieder die Gruppe Z_2^3;denn 
%S3 sind eigentlich die Perm von 123?
The lefthanded neutrino is taken to correspond to 
the unity transformation. All other fermion states within one 
family can be obtained from it by applying a nontrivial 
element of $S_s$.  

The elements of $D_4$ are identified as the lefthanded 
fermions of the first generation 
\begin{equation}
1(\nu_L), C_4^2(u_L^1), C_4(u_L^2), C_4^3(u_L^3),  
C_{2a}(e_L), C_{2b}(d_L^1), C_{2c}(d_L^2), C_{2d}(d_L^3) 
\label{eq7}
\end{equation}
Clearly, 
as yet this is not much more than a schematic representation 
of what we already know to be the content of one family 
but we shall see that it is part of a more natural and 
larger scheme which contains the fermions of all generations. 
Namely, one can extend the consideration from 1 to 3 families  
by going from the square to the cube (with 3 planes, 
cf. fig \ref{fig4}). The symmetry 
group of the cube is the octahedral group $O_h$. It is also 
of a direct product form $O_h\equiv P \oplus S_4$ with $S_4$ being the 
group of symmetry transformations of the tetrad, sometimes 
also called $T_h$ or $O$ (where the $O$ stands for the octahedron 
whose transformations it also describes). 
$O_h$ contains exactly the 48=2x24 elements needed for a one to one 
correspondence with the particles of the 3 families. 
The philosophy here is that by applying the 48 symmetry 
transformations of the cube on one of the 48 fermion states 
one can generate all the other 48 fermion states. 
This indicates that the fermions should have some 3 dimensional 
substructure which completely breaks the symmetry of the cubic.  
Such a symmetry breaking can be realized in a variety of different 
ways as will be shown now. 

\begin{figure}
\begin{center}
\epsfig{file=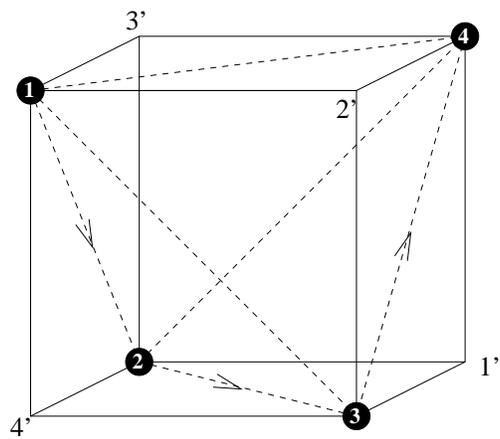,height=6.5cm,angle=270}
\vskip 0.8cm
\caption{The 'lefthanded' tetrad inside the cube. The path 
$1 \rightarrow 2 \rightarrow 3 \rightarrow 4$ corresponding 
to the identity is marked by arrows.  }
\label{fig5}
\end{center}
\end{figure}

The symmetry group $S_4$ of a tetrad is isomorphic 
to the group of permutations of 4 objects. This way the
24 symmetry tansformations on a tetrad can be viewed as the set of all
(directed, open) paths that connect 
the 4 corner points 1,2,3,4 of the tetrad.
From the tetrad a cube can be generated by applying 
the parity transformation. In fact there are two tetrads, 
a 'lefthanded' (with corner points 1,2,3,4) and a 'righthanded' 
(with corner points 1',2',3',4') embedded in a cube. 
(cf. fig. \ref{fig5}) related by P. 

From now on I will follow the philosophy that the spatial 
structure of a fermion (quark or lepton) is that of a tetrad.  
Furthermore, I assume that 
by applying the 24 symmetry transformations of the tetrad state 
one can create all 
24 fermion states of the three families out of one of these states. 
In order to guarantee that all of the symmetry transformations 
yield different states, the tetrad cannot be completely symmetric. 
For example, one may assume that there are 4 different preons  
sitting on the 4 corners of the tetrad. 
Another possibility is that the preons are identical, but the binding 
forces between them are different. In the following we shall 
pursue this latter option. More specifically, we shall 
assume that the bindings between the 4 preons are given 
according to the 24 permutations of the set {1,2,3,4}. 
One of the 24 permutations 
(namely the identity element) is visualized in fig. \ref{fig5} 
by the 3 arrows $1 \rightarrow 2 \rightarrow 3 \rightarrow 4$. 
Any other element $1234  \rightarrow abcd$ of $S_4$, denoted  
by $\overline{abcd}$ in the following, 
could be 
drawn as the path $a \rightarrow b \rightarrow c \rightarrow d$ 
in fig. \ref{fig5}.  

Starting with the 'lefthanded' tetrad,  
one can construct  
all the 24 lefthanded fermion states of the 3 generations.     
By applying the parity transformation 
$(x,y,z) \rightarrow (-x,-y,-z)$, righthanded fermions 
can be obtained. Any such righthanded state 
will be denoted by $\overline{a'b'c'd'}$ in the following. 
%As an example, $\overline{3'4'2'1'}$ 
%is shown in fig. \ref{fig6} 
%I have not undertaken to explicitly produce Dirac fermions out of 
%this construction. So admittedly, at the moment, only qualtitativ. 
As well known, a (Dirac) fermion $f$ has four degrees of freedom, 
of which only two, $f_L$ and $f_R$, have been described so far. 
The way to obtain antiparticles $\bar f_L$ and $\bar f_R$ is as follows: 
$\bar f_L$ is a righthanded object
and its preons should therefore form a righthanded tetrad 
$\overline{a'b'c'd'}$ with field values corresponding not to $f_R$ but 
to the complex conjugate of $f_L$. Similarly, 
$\bar f_R=\overline{abcd}$ with field values which are complex 
conjugate to $f_R$.  

As a side remark note that 
there might be something in the tetrad's centre, 
but this is not modified by $S_4$ nor parity transformations 
(cf. footnote 1 below).   

\begin{figure}
\begin{center}
\epsfig{file=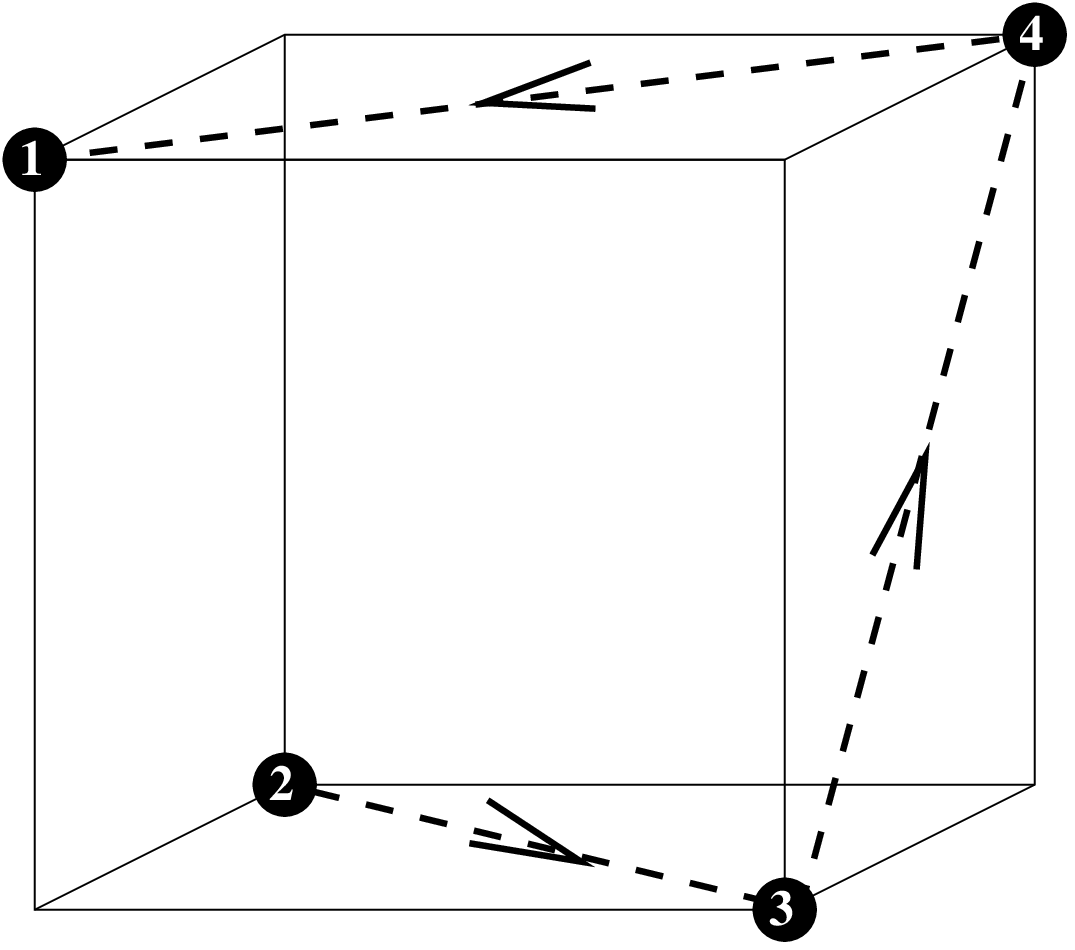,height=6.5cm,angle=270}
\vskip 0.8cm
\caption{ }
\label{fig55}
\end{center}
\end{figure}

\begin{figure}
\begin{center}
\epsfig{file=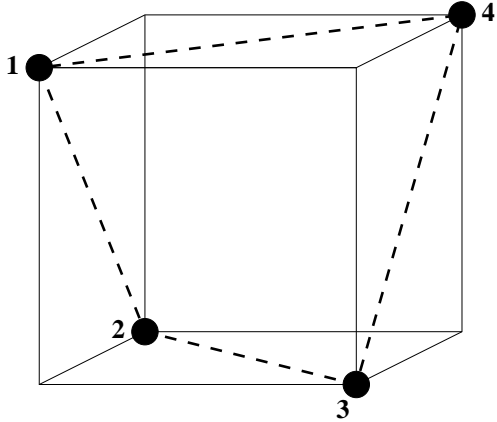,height=6.5cm,angle=270}
\vskip 0.8cm
\caption{Closed path I in $S_4$. The endpoints of the dashed lines 
form a tetrad. The projection of this path on a square 
leads to the equivalence of eqs. (\ref{eq7}) and (\ref{eq8}). }
\label{fig8}
\end{center}
\end{figure}

\begin{figure}
\begin{center}
\epsfig{file=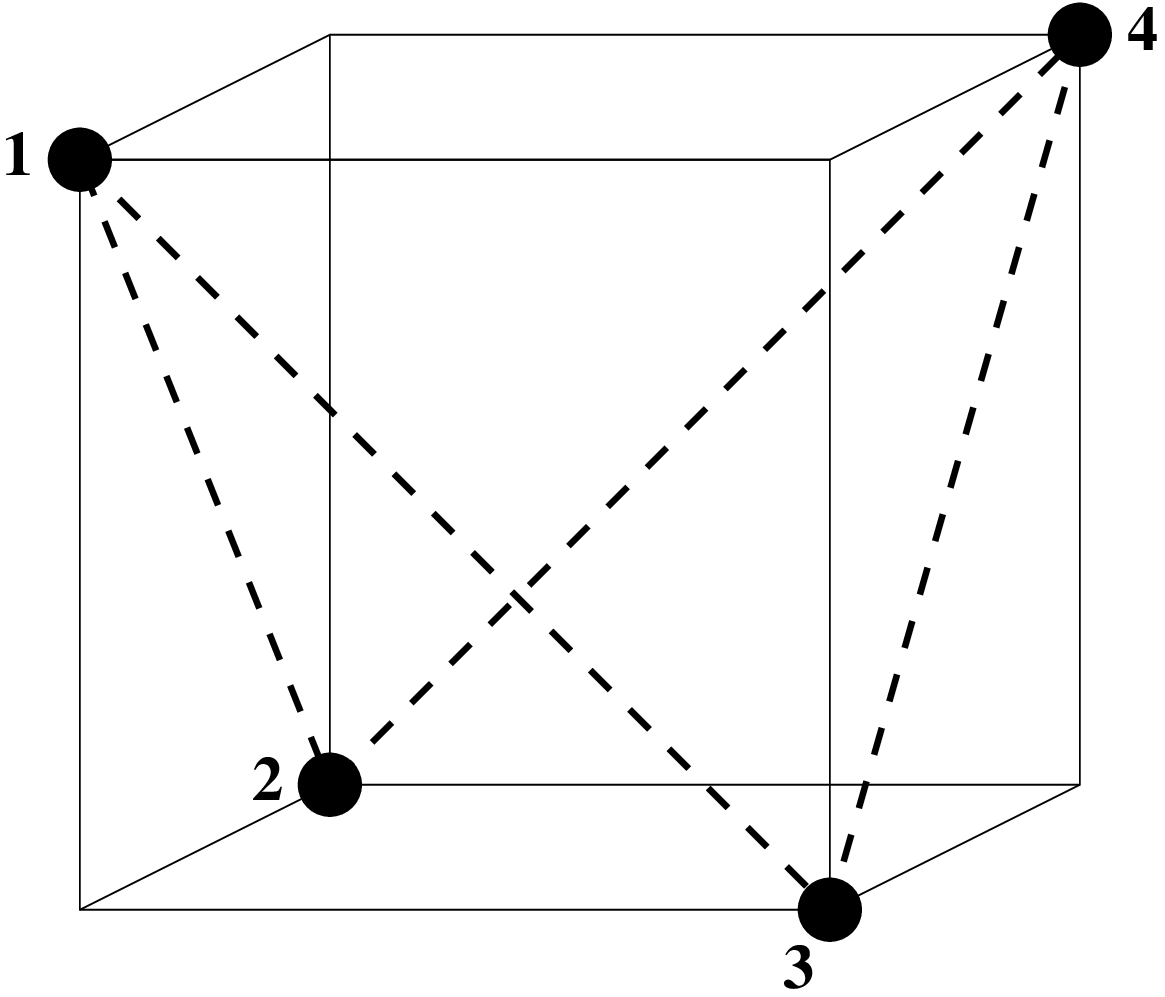,height=6.5cm,angle=270}
\vskip 0.8cm
\caption{closed path II }
\label{fig9}
\end{center}
\end{figure}

\begin{figure}
\begin{center}
\epsfig{file=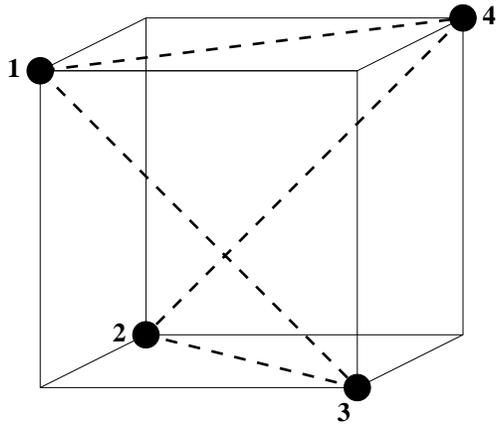,height=6.5cm,angle=270}
\vskip 0.8cm
\caption{closed path III }
\label{fig10}
\end{center}
\end{figure}

Let us now explicitly relate the elements of $S_4$ to the various 
members of the 3 generations. The geometrical model fig. \ref{fig5} 
naturally suggests a separation of the 24 permutations into 3 subsets. 
To see this, look at the figures \ref{fig8}, \ref{fig9} and \ref{fig10}, 
where the 3 possible closed paths which connects the points 1,2,3 and 4 
are shown. 
These closed paths consist of 4 links. 
Permutations lying on the path I (fig. \ref{fig8}) will be attributed 
to the members of the first family, permutations on path II (fig. \ref{fig9}) 
to the second family 
and path III (fig. \ref{fig10}) to the third family. 
For example, look at the lefthanded states of the first family   
\begin{eqnarray} \nonumber
 \nu_L &=& \overline{1234} \qquad u^1_L= \overline{2341} 
                          \qquad u^2_L= \overline{3412} \qquad 
                                     u^3_L= \overline{4123} \\
 e_L &=& \overline{4321} \qquad d^1_L= \overline{1432} 
         \qquad d^2_L= \overline{2143} \qquad 
                                     d^3_L= \overline{3214}
\label{eq8}
\end{eqnarray}
More precisely, these fermion states correspond to the various (open) paths 
consisting of 3 links which one can lay on the closed path fig. 
\ref{fig8}. A typical example of an open path 
(representing $u^1_L$) is shown in fig. \ref{fig55}. 

Using the assignments eq. (\ref{eq8}) 
one sees that a weak isospin transformation 
corresponds to reversing a permutation, i.e. reversing all 3 arrows 
in a figure like fig. \ref{fig55}. This way weak isospin is not 
any more a quantum number carried by a fundamental constituent 
but is determined by the binding of the state. 
Later on, a somewhat 
similar picture will be suggested for the understanding of 
electric and color charge of the first family. 
One can easily see that the assignments eqs. (\ref{eq7}) and 
(\ref{eq8}) are equivalent. One just has to project the closed path 
in fig. \ref{fig8} on a square. In so doing the permutation 
$\overline{2341}$ corresponds to the rotation $C_4^2$ etc.  

For completeness let us write down the $S_4$--assignment of the second  
and third generation. They correspond to the various paths consisting of 
3 links which one can draw into the closed loops depicted into 
figs. \ref{fig9} and \ref{fig10}.  
\begin{eqnarray} \nonumber
 \nu_{\mu L} &=& \overline{2134} 
       \qquad c^1_L= \overline{1342} \qquad c^2_L= \overline{3421} \qquad
                                     c^3_L= \overline{4213} \\
        \mu_L &=& \overline{4312} \qquad s^1_L= \overline{2431} 
                      \qquad s^2_L= \overline{1243} \qquad
                                     s^3_L= \overline{3124}
\label{eq9}
\end{eqnarray}
\begin{eqnarray} \nonumber
 \nu_{\tau L} &=& \overline{4231} \qquad 
       t^1_L= \overline{2314} \qquad t^2_L= \overline{3142} \qquad
                                     t^3_L= \overline{1423} \\
               \tau_L &=& \overline{1324} \qquad b^1_L= \overline{4132} 
                \qquad b^2_L= \overline{2413} \qquad
                                     b^3_L= \overline{3241}
\label{eq10}
\end{eqnarray}
It should be noted that there is an intimate connection 
between the 3 closed paths figs.   \ref{fig8}, \ref{fig9} and \ref{fig10} 
and the 3 planes in fig.  \ref{fig4}, i.e. the dimensionality of 
space. Fig. \ref{fig8} corresponds to the first family plane, 
Fig. \ref{fig9} to the second family plane and 
Fig. \ref{fig10} to the third family plane. 
This can be seen easily by drawing the octahedron with corners 
given by the middle points of the cube's face diagonals. 

Now for the question of electric and color charge. We want 
to adjoin quantum numbers to the various permutations above. 
There are several possibilities to solve this 
problem. As an example, we suggest the following construction. 
First of all, let us modify and refine the state identification given in 
eqs. (\ref{eq8}--\ref{eq10}) a little bit. Look at the four states 
\begin{equation}
 v_1 = \overline{4321} \qquad v_2= \overline{1432} 
          \qquad v_3= \overline{2143} \qquad
                                     v_4= \overline{3214} 
\label{eq11}
\end{equation}
which were preliminary identified as $e_L$, $d^1_L$, $d^2_L$ and $d^3_L$ 
in eq. (\ref{eq8}). Consider them as an orthonormal basis of an artificial 
vector space, i.e. $v_i \cdot v_j = \delta_{ij}$. 
Instead of $v_1$ we want to identify the linear combination 
${1 \over 4} (v_1 +v_2 +v_3 +v_4)$ as the lefthanded electron $e_L$, 
and the quark states $d^1_L$, $d^2_L$ and $d^3_L$ should span the 
subspace orthogonal to this linear combination, i.e. 
\begin{equation} 
e_L={1 \over 4} (v_1 +v_2 +v_3 +v_4)=(+++)_v
\label{eq12}
\end{equation}
\begin{equation}
d^1_L={1 \over 4} (v_1 -v_2 -v_3 +v_4)=(--+)_v
\label{eq13}
\end{equation}
\begin{equation}
d^2_L={1 \over 4} (v_1 +v_2 -v_3 -v_4)=(+--)_v
\label{eq14}
\end{equation}
\begin{equation} 
d^3_L={1 \over 4} (v_1 -v_2 +v_3 -v_4)=(-+-)_v
\label{eq15}
\end{equation}
where in the second part of these equations we have described 
the states by the relative sign of the coefficients of $v_i$ 
and $v_1$. Now we associate an additive quantum number 
$Q(+)_v=-{1 \over 3}$ and $Q(-)_v=0$ to those signs in order to obtain 
$Q(e_L)=-1$ and $Q(d_L)=-{1 \over 3}$. 

A similar construction can be carried out for the states 
$\overline{1234}$, $\overline{2341}$, $\overline{3412}$ and $\overline{4123}$, 
from which one can form linear combinations 
${1 \over 4}(\overline{1234}+\overline{2341}
+\overline{3412}+\overline{4123}) \equiv (+++)_w$ 
etc. in the same way as eqs.  (\ref{eq12}--\ref{eq15}). Note that this time the 
charge assignment should be $Q(-)_w=-{1 \over 3}$ and $Q(+)_w=0$  
in order to ensure $Q(\nu)=0$ and $Q(u)={2 \over 3}$. 

Now about the color charge: it is defined to operate 
on the three--dimensional space spanned by $d^1_L$, $d^2_L$ and $d^3_L$ 
(and similarly for the other quark flavors and helicities). 
A priori,  
the color degrees of freedom introduced here are purely real. 
However, one needs complex representations in order
to accomodate antiparticles. 
One possibility is  
that one artificially complexifies the vector 
spave spaned by $d^1_L$, $d^2_L$ and $d^3_L$. 
I think this is acceptable in the present situation, 
in which the true meaning of forming those linear 
combinations is unclear. It has to do with the specific 
way the preons on the sites of the 
tetrads are bound. 

Note that the construction presented here has 
a similarity to the charge assignment and the structure 
of the 'richon' model \cite{harari}. The states T and V 
of the Rishon model correspond to the various ways in which 
the above linear combinations are formed, $T \sim (+)$ and 
$V \sim (-)$. This is, however, the only similarity to preon 
models of that type. Those were constructed with 
an eye on obtaining a decent 
field theory but have some very unsatisfactory features. 
For example, 
there is no understanding whatsoever of parity violation. 
Parity violation and vector bosons will be discussed in the 
next section. 

In the literature there have been some attempts to use finite 
nonabelean groups to describe the family repetition structure. 
Usually, representations  are considered to 
model the generations, and a complicated Higgs sector is 
constructed to account for the observed mass differences 
between the fermions. In my opinion, this does not really 
solve the mass problem but just shifts it to another level. 
It would be much more desirable to understand the fermion 
masses dynamically, in the model at hand, for example, by 
understanding the differences between the bindings of preons 
1, 2, 3 and 4.  

\vskip2cm

{\bf 6. Vector Bosons}

Now that we have constructed all states of the fermion generations 
the most important question is how to understand their interactions. 
As is well known the interactions of fermions proceed through 
left-- and right--handed currents with the vector bosons, more 
precisely the lefthanded currents $\bar{F}_L \times f_L$ : $= 
\bar{F}_L \gamma_{\mu} f_L$ interact with the weak bosons 
and the sum of left- and righthanded currents interact with photons 
and gluons. The strength of the photonic and gluonic interaction 
is given by the electric and color charge, respectively. 

\begin{figure}
\begin{center}
\epsfig{file=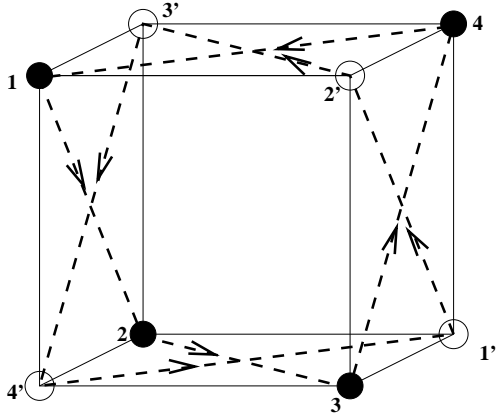,height=6.5cm,angle=270}
\vskip 0.8cm
\caption{A vector boson from fermion (full circle) 
and antifermion (open circle). It forms a closed loop 
in the first family plane.}
\label{fig11}
\end{center}
\end{figure}

The picture to be developed is that of 
the vector bosons as 
a sort of fermion--antifermion bound state. However, it will 
be constructed in such a way 
that the vector bosons do not 
'remember' the flavor of the fermion--antifermion pair 
from which they were 
originally formed. 
The way to obtain antiparticles (and Dirac fermions) is as follows: 
I have already shown that left-- and right--handed fields are 
interpreted as permutations of corners 1,2,3,4 and 1',2',3',4' 
in the cube (cf. fig. \ref{fig11}), for example 
$e_L=\overline{4321}$ and $e_R=\overline{4'3'2'1'}$. 
The antiparticle of a lefthanded fermion is a righthanded object
and its preons should form a righthanded tetrad.
Therefore, 
the $\bar e_L$ is defined to live on the righthanded tetrad 
($\overline{4'3'2'1'}$) but with field values corresponding 
to the complex conjugate of $e_L$. Similarly 
$\bar e_R=\overline{4321}$, with field values which are complex 
conjugate to $e_R$.  
Antifield configurations are denoted by open circles in 
fig. \ref{fig11}. 
\footnote{
I do not know whether the preons at the corners are 
real or complex, or whether one should prefer the bindings 
between them as the more fundamental objects.  
There are various disadvantages as to the existence of antipreons,  
both on 
the conceptual and on the explanation side. On the conceptual 
side the main disadvantage is that the preons themselves 
become more complicated than 
just real scalar pointlike particles without 
any further property than their simple 
superstrong interaction with 
neighbouring preons. 
On the explanation side I have found 
it difficult to accomodate parity violation -- everything 
is so unpleasantly P--symmetric in the pictures so far.   
An alternative one may follow is to do without antipreons, 
and to put the information of 
a quark or lepton being particle or antiparticle into the centre of 
the cube. More precisely, assume there is some nucleus $M_L$ at the centre 
of the lefthanded tetrads $ e_L, \nu_L, d^1_L, \dots$ 
and $\overline M_L$ at the centre of the righthanded tetrads  
$ \bar e_L, \bar \nu_L,\bar d^1_L, \dots$. 
Note that we do not assume the existence of a nucleus $M_R$ 
for the righthanded states $ e_R, \nu_R, d^1_R, \dots$ nor for their 
(lefthanded) antiparticles.  
We might assume its existence but 
for the sake of parity violation we must demand that $M_L$ and 
$M_R$ behave differently. In the following I shall 
assume for simplicity no $M_R$ at all.  
When a lefthanded current $\bar f_L \times f_L$ is formed, 
the $M_L$ and $\overline M_L$in the centre  
of the corresponding cube either annihilate or 
encircle each other. If they annihilate each other, a state 
is formed which cannot be distinguished from the corresponding 
right handed current $\bar f_R \times f_R$. That corresponds 
to the formation of a photon or a gluon. 
If they keep encircling each 
other, a Z or a W is formed, decaying very quickly after their short 
lifetime back to a fermion antifermion pair. 
The probability by which all these processes happen, is dictated 
by the various charges defined in section 5. A neutrino--antineutrino pair 
cannot form a photon nor a gluon, because it does not have an electric 
nor a color charge.}  
%I can think of two 
%ways to implement antiparticles in the present picture. 
%One possibility is to assume the existence of antipreons (white circles 
%in fig. \ref{fig11}) 
%in addition to the ordinary preons (black circles), so that 
%the corresponding antiparticle states can be formed. 
%The antiparticle of a lefthanded fermion is a righthanded object 
%and its preons should therefore form a righthanded tetrad. 
More precisely, 
the combinations of a lefthanded fermion of the first family 
and their righthanded 
antiparticles are shown in fig. \ref{fig11}. This way 
all the corners of the cube are filled. 
% by this combination. 
Fig. \ref{fig11} more or less represents how vector bosons 
should be imagined in this model. 

Fig. \ref{fig11} is a rather characteristic picture of a fermion--antifermion 
bound state. The point is that the vector boson interactions 
always take place within 
one family,  and fig. \ref{fig11} corresponds to interactions 
within the first family. 
One sees that the bindings between 
the links join together to form bindings along the plaquettes. 
Altogether, the bindings form an oriented closed circle of 
plaquettes. 
In the case of the second family interactions there is also a closed 
circle and it lies in the second family plane (cf. fig. \ref{fig4}) 
and similarly for interactions between members of the third 
family.  
The three planes can be rotated into each other to make the 
corresponding vector bosons identical. The difference to fermions 
will be understood 
better in a group representation approach to be discussed below. 

In my model, vector bosons are superpositions 
of fermion--antifermion states $\bar F \times f$ 
with the appropriate quantum numbers. 
The $\bar F \times f$ binding 
arises from interactions along 
the four body diagonals of the cube defined by fermions (1234) and 
antifermions (1'2'3'4'), i.e. interactions between the full and 
open circles in fig. \ref{fig11}. I shall come back to 
the body diagonals later. 

\begin{table}
\label{tab1}
\begin{center}
\begin{tabular}{|l|c|c|c|c|c|}
\hline
 & $I$ & $8C_3$ & $3C_4^2$ &
$6C_2$ & $6C_4$\\
 & Photon & Gluons & $W_{1,2,3}$ &
Leptoquarks X & Leptoquarks Y\\
\hline
$A_1$($U(1)_Y$) & 1 & 1  & 1  & 1  & 1 \\
$A_2$ & 1 & 1  & 1  & -1 & -1 \\
$E$ ($SU(2)_L$)   & 2 & -1 & 2  & 0  & 0 \\
$T_1$ & 3 & 0  & -1 & 1  & -1 \\
$T_2$($SU(3)_c$) & 3 & 0  & -1 & -1 & 1 \\
\hline
\end{tabular}
\bigskip
\caption{Character Table of $S_4$ and corresponding $SU(5)$ assignments. 
The form of the representation spaces are reminiscent of 
the $U(1) \times SU(2)_L \times SU(3)$ breaking structure 
from $SU(5)$.
%One may speculate that this correspondence has some real physical 
%significance, and that the real representation matrices of $S_4$ 
%become complexified to elements of $U(1) \times SU(2)_L \times SU(3)$ 
%by some yet unkown mechanism. 
%For the identity representation $A_1$ the character of each element is 1. 
%In $T_1$ the character of the identity is 3, while the characters of 
%$90^o$, $120^o$ and $180^o$ rotations are 1, 0 and -1 respectively. 
%With these initial characters  
%the characters of all other representations can be obtained 
%by means of the recursion relation 
%$\chi_{l+1}=\chi_l \chi_1 -\chi_l -\chi_{l-1}$. This relation 
%follows from remembering that the octahedral group can be 
%embedded in $SO(3)$ and in $SO(3)$ one has 
%$\bf{1} \otimes \bf{l}=(\bf{l+1})\oplus \bf{l} \oplus (\bf{l-1})$.   
}
\end{center}
\end{table}

For finite groups the number of irreducible representations (IR's) 
is equal to the number of conjugacy classes. In the present case 
the IR's are usually called $A_1$, $A_2$, $E$, $T_1$ and $T_2$ 
with dimensions 1, 1, 2, 3 and 3, respectively, and their 
characters are shown in table 1. $A_1$ is the identity 
representation. $A_2$ differs from $A_1$ by having a negative 
value for odd permutations. 
$T_1$ is the representation induced by 
the permutations of the corner points 1,2,3,4 of a tetrad 
in three dimensions. Its representation space is therefore 
the three dimensional space, in which the fermions live, 
i.e. {\it $T_1$ konstituiert den Anschauungsraum}. 
$T_2$ is obtained from $T_1$ by changing the sign of the representation 
matrices for the odd permutations. Finally, $E$ is induced by 
a representation of $S_3$ on the corners of a triangle,   
as discussed at the end of section 4 and fig. \ref{figt}, for example 
$E(\overline{2134})\bf{a}=\bf{b}$, $E(\overline{2134})\bf{b}=\bf{a}$, 
$E(\overline{2134})\bf{c}=\bf{c}=-\bf{a}-\bf{b}$, 
$E(\overline{1243})\bf{a}=\bf{b}$, $E(\overline{1243})\bf{b}=\bf{a}$, 
$E(\overline{1243})\bf{c}=\bf{c}=-\bf{a}-\bf{b}$, etc 
\cite{ham}.  

In order to obtain the vector bosons $\bar F \times f$, one should  
take the 9--dimensional product representation 
\begin{equation}
T_1 \times T_1 =A_1+E+T_1+T_2 
\label{eqzut}  
\end{equation}
On the right hand side, 
the term $T_1$ corresponds to arbitrary rotations of the closed loops 
of plaquettes, 
as claimed in connection with fig.  \ref{fig11}. 
%The interpretation of eq. (\ref{eqzut}) is as follows: 
$A_1$ corresponds to the photon, the totally symmetric singlet 
configuration, where all tetrad--antitetrad combinations 
contribute in the same way. 
$T_2$ is induced by 24 permutations 
of some objects I,II,III,IV (much like the $T_1$ on the left hand side of 
equation (\ref{eqzut}) was induced by   
the 24 permutations of 1,2,3,4). 
Finally, 
$E$ is induced by the 6 permutations on the triangle (fig. \ref{figt}). 

A possible interpretation of $T_2$ is as follows 
\footnote
{Alternatively, $T_2$ could represent 24 gluons 
which would then differ for the 3 families. The six 
permutations of the triangle might be 
the weak bosons $W^{1,2,3}$ and $W^{1,2,3}_{R}$.}
: By definition, 
the different vector bosons correspond 
to permutations of the cube's four body diagonals called I, II , III and IV, 
which define another group $S_4$.   
It is ordered not as in the case of fermions, equations 
(\ref{eq8})--(\ref{eq10}), but according to its conjugacy classes.  
In fact, 
%At the end of this section I want to point out that there is  
%a different possibility to understand the gauge bosons. Although I 
%did not find a relation to the model presented above, it is interesting 
%enough to be discussed. As already mentioned, the permutation group 
%$S_4$, the octahedral(=cubic) group $O$ and the tetrahedal group $T_d$ 
%are isomorphic \cite{ham}. 
the 24 elements of $S_4$ can be 
ordered in 5 conjugacy classes with 1, 3, 8, 6 and 6 elements. 
They are given as follows: 
\begin{itemize}
\item identity \\ 
      $\overline{I,II,III,IV}$ \\
      the U(1) gauge boson
\item 3 $C_2$ rotations by $\pi$ about the coordinates axes x, y and z \\  
      $\overline{II,I,IV,III}$, \qquad $\overline{III,IV,I,II}$, \qquad 
      $\overline{IV,III,II,I}$ \\  
      the SU(2) gauge bosons
\item 8 $C_3$ rotations by $\pm {2 \over 3}\pi$ about the cube diagonals  
        (like x=y=z)   \\ 
      $\overline{II,III,I,IV}$, \qquad $\overline{III,I,II,IV}$, \qquad 
      $\overline{II,IV,III,I}$, \qquad  
      $\overline{IV,I,III,II}$, \\ 
      $\overline{III,II,IV,I}$, \qquad
      $\overline{IV,II,I,III}$, \qquad
      $\overline{I,III,IV,II}$, \qquad $\overline{I,IV,II,III}$ \\
      the gluons
\item 6 $C_4$ rotations by $\pm {\pi \over 2}$ about the coordinate axes 
      \\ 
      $\overline{II,I,III,IV}$, \qquad $\overline{III,II,I,IV}$, 
      \qquad $\overline{IV,II,III,I}$,  \\
      $\overline{I,III,II,IV}$, \qquad $\overline{I,IV,III,II}$, 
      \qquad $\overline{I,II,IV,III}$ \\ 
      leptoquarks 
\item 6 $C_2 '$ rotations by $\pi$ about axes parallel to the face 
      diagonals (like x=y, z=0) \\ 
      $\overline{II,III,IV,I}$, \qquad $\overline{II,IV,I,III}$, 
      \qquad $\overline{III,IV,II,I}$, \\
      $\overline{III,I,IV,II}$, \qquad $\overline{IV,III,I,II}$, 
      \qquad $\overline{IV,I,II,III}$ \\
      leptoquarks 
\end{itemize} 
where reference is made to the cartesian coordinates x,y and z with 
origin at the cube's centre. 
This ordering is reminiscent of the ordering of gauge bosons in 
grand unified theories where there are leptoquarks in addition 
to the 8 gluons and the four electroweak gauge fields. 
The elements of the first two classes form Klein's 4--group 
(an abelean subgroup of $S_4$), 
whereas the elements of the first three classes form the 
nonabelean group of even permutations. 

In summary, the interpretation of eq. (\ref{eqzut}) is as follows:
As discussed before, the fermions constitute ordinary three--dimensional 
space. As soon as two fermions approach each other to form 
a vector boson, space opens up to 9 dimensions. Three of them 
correspond to ordinary space, whereas the remaining six decompose 
into 1+2+3 dimensional representation spaces $A_1$, $E$ and $T_2$ of $S_4$. 
They become fibers to ordinary space. It remains to be shown how the complex 
structure of a $U(1)\times SU(2)\times SU(3)$ Lie algebra 
arises. 

Since parity violation is not present in these pictures, 
I want to add an alternative related to the observation 
that there are two 1--dimensional and two 3--dimensional, 
but only one 2--dimensional IR of $S_4$. 
One could relate parity transformations to even--odd 
transitions between permutations by modifying the assignments 
made in equations 
(\ref{eq8})--(\ref{eq10}), namely 
\begin{eqnarray} \nonumber
 \nu_L &=& \overline{1234} \qquad u^1_R= \overline{2341}
                          \qquad u^2_L= \overline{3412} \qquad
                                     u^3_R= \overline{4123} \\
 e_L &=& \overline{4321} \qquad d^1_R= \overline{1432} 
                  \qquad d^2_L= \overline{2143} \qquad
                                     d^3_R= \overline{3214}
\label{eq8s}
\end{eqnarray}
i.e. assigning odd permutations to righthanded states. 
According to the character table 1 the character of $E$ vanishes 
for odd permutations. Therefore, there is no action of $E$ on 
righthanded fermions. In contrast, the products $A_1\times A_2$ 
and $T_1\times T_2$ act like $-\gamma_5$ on left and 
righthanded fermions.

\vskip2cm

{\bf 8. Conclusions}

According to present ideas the elementary particles (leptons, 
quarks and vector bosons) are pointlike and their mathematical 
description follows this philosophy (Dirac theory, Yang--Mills theory). 
They certainly receive an effective extension by means of quantum 
effects, but these are fluctuations and do not affect the primary 
idea of pointlike objects. 

In contrast,
in the preon picture the observed fermions naturally have an 
extension right from the beginning.  
This seems to be difficult to accomodate because their radius should 
be of the order of their inverse masses.  Following t'Hooft 
one may assume that there is a symmetry principle which leaves 
the masses small. 

The models in this paper do not 
allow to make quantitative predictions of fermion masses. 
Some qualitative statements about fermion masses can be found 
in sections 2 and 3.   
As compared to the binding energy, all fermion masses 
(including $m_t$)  
are tiny perturbations which might be induced by some radiative mechanism 
of  the 'effective'
standard model interactions  
leading to masses $\sim \alpha^F$ for the F--th family.  
The 'textures' of those masses have been discussed in 
section 2 (cf. \cite{rrr} for more elaborate approaches). 

Within the models of sections 4, 5 and 6 one may assume that 
the fermions are basically massless by some symmetry and 
that there are small symmetry breaking effects within the 
family planes  
leading to different family masses eq. \ref{eq2}. 
%For example, 
%there might be a gradient g(x), g(y) and g(z) 
%along the x--, y-- and z--direction 
%such that the family loops figs. \ref{fig8}--\ref{fig10} get 
%different masses   
%\begin{equation}
%m_I \sim g(x)^2 \quad m_{II} \sim g(y)^2 \quad m_{III} \sim g(z)^2 
%\label{eqm1}
%\end{equation}
%with the constant of proportionality being given by 
%$C=({g(x)g(y)g(z) \over 1000 MeV^{3/2}})^2$ and $g(x)^2=1$ MeV, 
%$g(y)^2=100$ MeV $g(x)^2=10$ GeV. This way the family masses 
%are essentially determined by the links that connect the preons. 
%One should remember that the same holds true for 
%the construction of charges in our model (cf. section 5). 

Whatever this symmetry principle may be, there is still the question how 
large the radius R of the quarks and leptons is. In principle, there 
are three possibilities, it may be large ($\sim 1$ TeV$^{-1}$), small 
($\sim M_{Planck}^{-1}$) or somewhere in between. In the first case 
there will be experimental signals for compositeness very soon. 
In the second case there will never be direct experimental 
indications and it will be difficult to verify the preon idea. 
Furthermore, in that case one would have the GUT theories as correct 
effective theories whose particle content would have to be 
explained. In addition, it may be necessary to modify the theory 
of relativity. In fact, the superstring models are a realization 
of this idea, the 'preons' being strings instead of point particles. 

Personally, I like the scenario $R \sim M_{Planck}^{-1}$ reasonably well. 
In the model presented in this paper, the preons are pointlike and 
sitting on a cubic lattice. This lattice would have to fluctuate in 
some sense to reconstitute Lorentz invariance. This certainly 
raises many questions which go beyond the scope of this article. 
For example, the renormalization of gravity would be modified 
because high energies ($> M_{Planck}$) would be cut away by the 
lattice spacing. 

As for the third possibility 1 TeV $<< R << M_{Planck}$, 
gravity and its problems 
play no role and my models are just a more or less consistent 
picture of particle physics phenomena. Since no attempt was  
made to explicitly construct the states
of quarks and leptons in their known complex
representations they are at best a qualitative guideline for 
understanding. 
%a littel bit of UNDERSTANDING substraucture
%No predictive power, because no quantitative model. 
%Even if one follows the philosophy suggested in 
%this paper, there are certainly better solutions 
%in the way how to implement the basic phenomenological 
%facts, like antiparticles, partity violation etc. 
%In this article I merely wanted to show that 
%in this framework solutions exist. 
I did not write a Lagrangian for the preons and just speculated about 
their interactions. The ultimate aim would be to construct a Lagrangian 
and derive from it an effective interaction between Dirac fermions 
and gauge fields.  

\vskip2cm

{\bf Acknowledgements}
 
With this paper my scientific efforts come to an end.
After 18 years of hard work I have not been able to find
a reasonable position in physics. 
%Particle phenomenology
%in Germany is nowadays dominated by career opportunists 
%which effectively prevent me
%to continue my beloved work.

\vskip2cm

%\newpage 

%\newpage

\end{document}